\newif\iffigs\figstrue
\iffigs
   \documentstyle[12pt,epsf]{article}
\else
   \documentstyle[12pt]{article}
\fi

\begin{document}
\def\bx{{\bf x}}
\def\bZ{{\bf Z}}
\def\dop#1{{\rm d}\hskip -1pt #1}
\def\real{{\rm Re}\hskip 1pt}
\def\ee#1{{\rm e}^{#1}}
\def\trace{{\rm Tr}\hskip 1pt}
\def\diag{{\rm diag}}
\def\ii{{\rm i}}
\def\cJ{{\cal J}}
\def\cI{{\cal I}}
\def\cF{{\cal F}}
\def\Udag{U^{\dagger}}
\def\Vdag{V^{\dagger}}
\def\duepisuL{{2\pi\over L}}
\def\ev#1{\langle #1 \rangle}
\newcommand{\NP}[1]{Nucl.\ Phys.\ {\bf #1}}
\newcommand{\PL}[1]{Phys.\ Lett.\ {\bf #1}}
\newcommand{\NC}[1]{Nuovo Cim.\ {\bf #1}}
\newcommand{\CMP}[1]{Comm.\ Math.\ Phys.\ {\bf #1}}
\newcommand{\PR}[1]{Phys.\ Rev.\ {\bf #1}}
\newcommand{\PRL}[1]{Phys.\ Rev.\ Lett.\ {\bf #1}}
\newcommand{\MPL}[1]{Mod.\ Phys.\ Lett.\ {\bf #1}}
\newcommand{\IJMP}[1]{Int.\ J.\ Mod.\ Phys.\ {\bf #1}}
\newcommand{\JETP}[1]{Sov.\ Phys.\ JETP {\bf #1}}
\newcommand{\TMP}[1]{Teor.\ Mat.\ Fiz.\ {\bf #1}}
\renewcommand{\topfraction}{.8}
\renewcommand{\textfraction}{.2}
\renewcommand{\arraystretch}{1.3}
\arrayrulewidth 0.3pt 
\doublerulesep 0.7pt
\def\thefootnote{\alph{footnote}}
\def\mycaptionl#1#2{%
\begin{center}
\hskip 1pt\vskip -1cm
\begin{minipage}{14cm}
\small {\bf #1}: {\sl #2}
\end{minipage}
\null\hskip 1pt\vskip -0.2cm
\end{center}}
\def\mytcaptionl#1#2{%
\begin{center}
\begin{minipage}{14cm}
\small {\bf #1}: {\sl #2}
\end{minipage}
\null\hskip 1pt\vskip -0.3cm
\end{center}}
\def\mycaptions#1#2{%
\hfil{\small {\bf #1}: {\sl #2}} \hfil}%
\hskip 10cm \vbox{\hbox{DFTT 30/96}\hbox{NORDITA 96/37P}}
\vskip 1cm
\begin{center}
\Large{\bf A solvable twisted one-plaquette model}
\end{center}
\vskip 0.7cm
\begin{center}
{\large M. Bill\'{o}\,\footnote{Supported by the Istituto Nazionale 
di Fisica Nucleare, Italy}
\footnote{e-mail: {\tt billo@alf.nbi.dk}} }
\vskip 0.1cm
{\small\it Nordita, Blegdamsvej 17, Copenhagen \O, Denmark}
\vskip 0.6cm
{\large A. D'Adda\,\footnote{e-mail: {\tt dadda@to.infn.it}}}
\vskip 0.1cm
{\small\it Istituto Nazionale di Fisica Nucleare, Sezione di Torino and
Dipartimento di Fisica Teorica dell'Universit\`a di Torino,
via P.Giuria 1, I-10125 Turin,Italy}
\end{center}
\vskip 0.7cm
\begin{center}
{\bf Abstract}\\
\vskip 0.4cm
\begin{minipage}{14cm}
\small 
We solve a hot twisted Eguchi-Kawai model with only timelike plaquettes
in the deconfined phase, by computing the quadratic  quantum fluctuations 
around the classical vacuum. The solution of the model has some novel features:
the eigenvalues  of the time-like link variable are separated in $L$ bunches,
if $L$ is the number of links of the original lattice in the time direction,
and each bunch obeys a Wigner semicircular distribution of eigenvalues. 
This solution becomes unstable at a critical 
value of the coupling constant, where it is argued that a condensation of 
classical solutions takes place. This can be inferred by comparison 
with the heat-kernel model in the hamiltonian limit, and the related 
Douglas-Kazakov phase transition in QCD2.
As a byproduct of our solution,  we can reproduce the dependence of the 
coupling constant from the parameter describing the asymmetry of the lattice, 
in agreement with previous results by Karsch.
\end{minipage}
\end{center}
\vskip 0.9cm
\setcounter{footnote}{0}
\def\thefootnote{\arabic{footnote}}
\section{Introduction}
The twisted one-plaquette models, or Eguchi-Kawai models \cite{EK,das}, 
have been extensively 
studied in the last fifteen years for their remarkable property of 
satisfying, in the large $N$ limit, the same loop equations as the 
corresponding lattice gauge theories. 
The reduction of the space-time degrees of freedom to
internal degrees of freedom is the most startling property of these models.
Their solution  would amount to finding 
the master field for lattice gauge theories, 
and to solving them in the large $N$ limit. 
Unfortunately, in spite of the reduction to one plaquette,
the saddle point structure of the Eguchi-Kawai model in quite complicated, 
due to the effect of the twists, and a complete solution appears to be 
beyond the reach of our present techniques.

In this paper we study a  twisted one-plaquette model, which is much
simpler than the complete twisted Eguchi-Kawai model, but whose
solution can  be explicitly found, at least in the weak coupling
(or high temperature) regime, in terms of eigenvalue distribution.
The model under consideration is defined by the partition function
\begin{equation}
\label{model}
Z = \int \hskip -2pt DV \int\hskip -1pt \prod_{\mu = 1}^d DU_{\mu}\,\, 
\ee{\beta N \real 
\ee{\ii{2 \pi \over L}} \sum_{\mu = 1}^d \trace 
(V U_{\mu} \Vdag \Udag_{\mu} )}
\end{equation}
and in the large $N$ limit it is equivalent, as shown later in the paper,
to a $d+1$ dimensional lattice gauge theory at
finite temperature, with $L$ links in the compactified time dimension, and
with only the couplings induced by time-like plaquettes.
Hence the action described by eq. (\ref{model}) arises naturally in finite
temperature lattice gauge theories as a zero order approximation 
in which space-like plaquettes are completely neglected.
This approximation has already been discussed in
previous papers \cite{largeN,bcdp} and it will be reviewed in the following 
section.

The main result of our paper consists in solving the model in the broken 
phase,  which occurs for values of $\beta$ larger than a critical value
$\beta_{\rm cr}$ and is characterised by a non vanishing vacuum expectation 
value of the Polyakov loop ${\cal P} =\trace V^L$. 
The solution has some novel, interesting features, namely an  
eigenvalue distribution for the matrix $V$ which consists of $L$ separated 
bunches  centered around the $L^{\rm th}$ roots of unity.  
The eigenvalue distribution within each bunch is given, with very good
approximation, by a semicircular Wigner distribution of calculable radius.
The splitting of the eigenvalue distribution into several bunches is a new
feature that makes the model in consideration interesting in itself 
and susceptible of applications in other branches of physics, 
possibly for instance condensed matter physics. 

The paper is organised as follows.
In section \ref{reduction} the twisted
Eguchi-Kawai reduction in $3+1$ dimensions is considered in the
framework of finite-temperature lattice gauge theories, and a scheme of
dimensional reduction is proposed that allows to extend the analysis to
$d+1$ dimensions.
In section \ref{quadmodel} we derive the effective model for the Polyakov loops
in the deconfined phase, up to quadratic order in the fluctuations.
In section \ref{solution} the solution of the effective model and the phase 
transition occurring at a critical value of the coupling are discussed 
for a model with arbitrary number $L$ of lattice 
links in the time direction. 
In this context we also derive the rescaling of the coupling $\beta$
with the parameter $\rho$, describing the asymmetry of the original lattice,
and we find results in excellent agreement with the ones by Karsch in 
~\cite{Karsch}.
In section \ref{conclsec} we present our conclusions.
\section{Dimensional reduction in finite temperature large $N$ LGT}
\label{reduction}
\subsection{Twisted reduction in $3+1$ dimensions}
The symmetric twisted Eguchi-Kawai model (TEK ) in 3 + 1 dimensions is a one 
plaquette model which is equivalent, in the large $N$ limit, to a lattice 
SU(N) pure gauge theory defined in a periodic box of size $L$ , with $N=L^2$.
Its action is given by\footnote{In this formula $\mu,\nu$ can assume the 
values $0,1,2,3$. Afterwards we will always utilise these indices to
enumerate the space dimensions only.}:
\begin{equation}
\label{symmek}
S_{\rm sTEK} = N \beta \sum_{\mu > \nu} \real \trace 
(\ee{\ii {2 \pi \over L}} U_{\mu} U_{\nu} \Udag_{\mu} \Udag_{\nu} )~~.
\end{equation}
In the large $N$ limit the size of the box goes to infinity in all 
directions.
If we want to describe a finite temperature gauge theory we need a lattice 
whose size in the compactified time direction remains finite 
also in the large $N$ limit. 
This presents a problem in the TEK approach where the size
of the box is linked to $N$. The problem can be overcome in two 
different ways.
We can define the theory in an asymmetric box, whose spatial extent goes
to infinity in the large $N$ limit while  its extent in the time direction
remains finite. Several of such boxes with the corresponding twists exist in
the literature \cite{klvb,fabr}.
On the other hand, we can keep a symmetric box of size $L$ but with different
lattice spacings $a$ and $a_{\tau}$ respectively in space and time directions.
If $T$ is the temperature, then $1/T$ is the size of compactified time 
dimension, given by:
\begin{equation}
\label{temp}
 {1 \over T} = L a_{\tau}~~.
\end{equation}
Let us introduce the asymmetry parameter $\rho = a/a_{\tau}$. 
Then from (\ref{temp}) we have
\begin{equation}
\label{asy}
\rho = a T L~~.
\end{equation}
{}From eq. (\ref{asy}) it is clear that in order to keep $T$ finite 
the parameter
$\rho$ has to go to infinity like $L=N^{1/2}$ in the large $N$ limit. 
The introduction of different lattice spacings is equivalent to the
introduction of different couplings $\beta_s$ and $\beta_t$ for
space-like and time-like plaquettes respectively. The relation between
these new couplings and the asymmetry parameter $\rho$  is given by
\begin{equation}
\label{asymmetry}
\beta_t(\rho) =  \rho( \beta_{\rm symm} + 4 c_{\tau}(\rho))
\hskip 0.8cm ; \hskip 0.8cm
\beta_s(\rho) = {(\beta_{\rm symm} + 4 c_{\sigma}(\rho)) \over \rho}~~,
\end{equation}
where $\beta_{\rm symm}$ is the coupling corresponding 
to a symmetric lattice, namely to $\rho =1$. 
The terms proportional to $\beta_{\rm symm}$ in (\ref{asymmetry}) stem from
the requirement that a Yang-Mills theory is reproduced 
in the naive or classical
continuum limit. The functions $c_{\tau}$ and $c_{\sigma}$ on the other hand
represent the result of quantum corrections and have been evaluated by 
F. Karsch in \cite{Karsch}. This point will be discussed more in 
detail in the section \ref{qresc}.

The resulting one-plaquette model in this approach is then described by the
following partition function:
\begin{eqnarray}
\label{hotEK}
Z_{\rm hTEK} & = & \int\hskip -2pt DV \int\hskip -1pt \prod_{\mu=1}^3 DU_\mu\,\,
\exp \left(S_{\rm hTEK}\right) \nonumber\\
S_{\rm hTEK} & = &  
N \beta_t(\rho) \sum_{\mu=1}^3 \real \trace (\ee{\ii {2 \pi \over L}
} V U_{\mu}\Vdag \Udag_{\mu} )
+ N \beta_s(\rho) \sum_{\mu>\nu=1}^3 \real \trace 
(\ee{\ii {2 \pi \over L}} U_{\mu} U_{\nu} \Udag_{\mu} \Udag_{\nu} )\,
\nonumber\\
\end{eqnarray}
where we have denoted with $V$ the link variable in the time direction, and 
with $U_{\mu}$ the ones in space directions. It is important to remember that
only traces corresponding to closed loops in the 
original theory\footnote{That is, the $d+1$-dimensional LGT (defined on a box
of size $L$) from which the model (\ref{symmek}) [or the model (\ref{hotEK})] 
is obtained by eliminating the space-time dependence 
of the gauge fields via the ``twisted Eguchi--Kawai'' reduction 
(see \cite{das}, and similar arguments in next section \ref{twistred}).} 
have physical meaning, like for instance the Polyakov loop 
which is given by
\begin{equation}
\label{polya}
{\cal P} = {1 \over N} \trace V^L~~.
\end{equation}
We shall not consider here in detail the models based on the use of 
asymmetric boxes, which have the advantage of having $\beta_s=\beta_t=\beta$ 
but the disadvantage of far more complicated twists, at least for the
space-like plaquettes.
It should be noticed however that, in the approximation where 
space-like plaquettes are neglected, the model based on asymmetric 
boxes leads to the same action  given in (\ref{model}).
\subsection{The perturbative expansion in $\beta_s$}
If we were able to perform the exact integration over the space-like link 
variables $U_{\mu}$ in eq. (\ref{hotEK}), we would obtain
the exact effective action for $V$, and thus for the Polyakov loop,
in the large $N$ limit.
However this is  too difficult, and  we shall follow a
perturbative approach consisting in an expansion of the partition
function (\ref{hotEK}) in powers of $\beta_s$.  
In the present paper we shall consider only the zeroth order of 
this expansion, namely  the partition function (\ref{model}) 
which corresponds to neglecting the space-like
plaquettes altogether. An explicit calculation of the first non
trivial order in $\beta_s$ seems to be within reach of the techniques
developed in the following section, but it will be left to
future investigation. The extent to which the zero order approximation is 
reliable was discussed in ref.~\cite{largeN}, where the same approximation
was used to study, in the large $N$ limit, the effective 
action for the Polyakov 
loop in a lattice gauge theory with both a Wilson and a heat kernel action.
The difference between the approach of the present paper and the one
of ref.~\cite{largeN} is that the reduction to a one plaquette model
was obtained there, for the Wilson action, by an approximate Migdal-Kadanoff
\cite{kad} bond-moving scheme.
Instead our starting point eq. (\ref{hotEK}) is exact.

The perturbative expansion in $\beta_s$ was studied in even more details
in ref.~\cite{bcdp}, where the expansion was carried out, for the $SU(2)$
lattice gauge theory, up to the order $\beta_s^2$.
The feature that emerges from ref.s \cite{largeN,bcdp} is that
the zeroth order approximation,
although it correctly predicts the existence of a deconfinement transition
for large temperature, fails to reproduce the deconfinement temperature in
$3+1$ dimensions for lattice sizes\footnote{The lattice size $n_t$
introduced here corresponds to the number of links in the
time direction in a lattice with the same lattice spacing in the space 
and time directions. So it is related to the lattice size $L$
of the Eguchi-Kawai reduction of eq.(\ref{hotEK}) by the relation
$ n_t = {L \over \rho}$.} $n_t$ in the time direction
greater than 2.
This is related to the fact that in the zeroth order approximation the
critical value of $\beta_t$ rescales linearly with $n_t$ rather than
with the logarithmic law required by asymptotic freedom. A correct
prediction of the latter behaviour would presumably require an exact
derivation, namely to all orders in $\beta_s$, of the effective action 
for the Polyakov loop; however in the $SU(2)$ theory the first non 
trivial order in $\beta_s$ already provides a striking improvement 
in the scaling properties ~\cite{bcdp}.
We expect the same to happen in the large $N$ case, but in this paper
we shall restrict ourselves, as already mentioned, to the zeroth order
approximation, the emphasis being more  on the  the matrix model 
(\ref{model}) itself and its interesting features.
It should be noticed also that, unlike the $3+1$ dimensional case,
in $2+1$ dimensions the full theory has the same linear rescaling with
$n_t$ as the $\beta_s=0$ approximation, and that there are indications that
such approximation is in that case reliable also for large $n_t$.
\subsection{Twisted reduction in $d+1$ dimensions. Weak coupling vacuum}
\label{twistred}
We have so far  considered the reduction to a one plaquette model of  
finite temperature lattice gauge theory only in $3+1$ dimensions. 
We want now to show here
that the partition function given in (\ref{model}) can be obtained
as the one plaquette reduction of the large $N$ finite temperature
lattice gauge theory in $d+1$ dimensions, in the zeroth order approximation
$\beta_s=0$.
For even space-time dimensions we could prove this by using the existing
twisted Eguchi-Kawai models, essentially in the same way 
as in $3+1$ dimensions.
However in odd space-time dimensions consistent twists have not been 
constructed, and we have therefore to rely on a different type of
approach.  

We consider a lattice gauge theory defined on a lattice
which is infinite in the space directions and of length $L$ with periodic
boundary conditions in the time direction. The strategy is to reduce the 
height of the lattice in the time direction to one by using the same type 
of approach as in the Eguchi-Kawai model, and then to eliminate the
dependence from space coordinates by assuming that a constant master
field dominates in the large $N$ limit.
Let us start with the action
\begin{eqnarray}
\label{finitet}
S(L)& =& N \sum_{\vec x} \sum_{t=1}^L  
\real \trace\left(\beta_t  \sum_{\mu=1}^{d} 
V(\vec x,t) U_{\mu}(\vec x,t+1) \Vdag (\vec x + \hat \mu, t) \Udag_{\mu}
(\vec x,t) \right.\nonumber \\ 
& + & \left. \beta_s \sum_{\mu > \nu=1}^d U_{\mu}(\vec x,t) 
U_{\nu}(\vec x+\hat \mu,t)
\Udag_{\mu}(\vec x +\hat \nu,t) \Udag_{\nu}(\vec x,t)\right)~~.
\end{eqnarray}
Notice that in (\ref{finitet}) we kept two independent couplings
$\beta_s$ and $ \beta_t$. Unlike the case of eq. (\ref{hotEK}), 
however, this is not necessary in order to be able to take the 
limit $N\rightarrow \infty$, because $L$ and $N$ are now unrelated,
and we  will be free to choose $\beta_s=\beta_t$ at our convenience. 
The naive prescription for the reduction of the time degrees of freedom
would be
\begin{eqnarray}
\label{ttekred}
V(\vec x,t) & \rightarrow V(\vec x)~~, \nonumber\\
U_{\mu}(\vec x,t) & \rightarrow U_{\mu}(\vec x)~~.
\end{eqnarray}
It is easy to show by  standard methods (see for instance~\cite{das} )
that the resulting action, which can be identified with $S(1)$, 
leads to the same set of loop equations in the large $N$ limit as the full 
$S(L)$ theory, {\it provided all loops which are closed in the reduced
lattice ($L=1$) but correspond to open loops in the original lattice
have vanishing expectation value}.
Just as in the untwisted Eguchi-Kawai model, this would be
granted by the symmetry
\begin{equation}
\label{symm}
V(\vec x) \rightarrow \ee{i{2 \pi n \over N}} V(\vec x)
\end{equation}
which is satisfied by $S(1)$. The trace along ``open" lines is not
invariant under the symmetry (\ref{symm}) as they do not contain the
same number of $V(\vec x)$ and $\Vdag(\vec x)$ fields. So these 
contributions vanish  unless the symmetry is broken.
The symmetry however is actually broken in the  weak
coupling regime, that is also in the continuum limit, where $V(\vec x)$
is close to ${\bf 1}$ (more generically to an element of ${\bf Z}_N$)
and the traces of open lines do not vanish.
Consequently the reduction prescription must be endowed with twists,
namely it must be of the type
\begin{eqnarray}
\label{tekred}
V(\vec x,t) &\rightarrow& D(\vec x,t)V(\vec x) D^{\dagger}(\vec x,t)~~,
\nonumber\\ U_{\mu}(\vec x,t) &\rightarrow & D(\vec x,t)U_{\mu}(\vec x) 
D^{\dagger}(\vec x,t)~~, 
\end{eqnarray}
where $D(\vec x,t)$ is given by
\begin{equation}
\label{red}
D(\vec x,t) = (\Gamma)^{\sum x_i} (\Gamma_0)^t~~,
\end{equation}
with $\Gamma$ and $\Gamma_0$ traceless $SU(N)$ matrices 
satisfying the 't Hooft algebra
\begin{equation}
\label{hooft}
\Gamma \Gamma_0 = \ee{\ii {2 \pi \over N} m} \Gamma_0  \Gamma ~~.
\end{equation}
In the last equation $m$ is an integer number to be determined.
By performing in (\ref{finitet}) the  replacement (\ref{tekred}) and  
redefining the variables according to the substitution $U_{\mu}(\vec x) 
\rightarrow 
U_{\mu}(\vec x) \Gamma$ and $V(\vec x) \rightarrow V(\vec x) \Gamma_0$
we obtain
\begin{eqnarray}
\label{ttekaction}
Z_R & = & \int\prod_{\vec x}\bigl[DV(\vec x)\prod_{\mu=1}^d 
DU_{\mu}(\vec x)\bigr]\, \exp (S_R)~~,
\nonumber\\
S_R &=& \beta_t N \sum_{\vec x} \sum_{\mu=1}^d  
\real~ \ee{\ii{2\pi m\over N}}\trace
\bigl[U_{\mu}(\vec x) V(\vec x + \hat \mu)  
\Udag_{\mu}(\vec x) \Vdag(\vec x)\bigr] +
\nonumber\\
&&\beta_s N \sum_{\vec x} \sum_{\mu > \nu} \real\trace \bigl[ 
U_{\mu}(\vec x) U_{\nu}(\vec x+ \hat \mu) \Udag_{\mu}(\vec x+ \hat \nu) 
\Udag_{\nu}(\vec x)\bigr]~~.
\end{eqnarray} 
Notice that, unlike the twisted Eguchi-Kawai model, the twists are 
present in (\ref{ttekaction}) only in the contributions from time-like 
plaquettes, as the  reduction has been done only in the time direction.

Consider now the loop equations for the reduced theory (\ref{ttekaction}).
We already remarked that as long as the symmetry (\ref{symm}) is unbroken 
the loop equations of the reduced theory coincide with the ones of 
(\ref{finitet}). We will show now that in the twisted theory this is the
case {\it also in the weak coupling regime}. 
Indeed in the extremely weak coupling the fields tend to their vacuum
configurations,  which in our twisted reduced theory are
\begin{eqnarray}
\label{ttekvacuum}
U_i(\vec x) &\rightarrow  &P_{{N \over m}}\otimes {\bf 1}_{m}~~,\nonumber\\
V(\vec x) &\rightarrow &Q_{{N \over m}}\otimes {\bf 1}_{m}~~,
\end{eqnarray}
where ${\bf 1}_{m}$ is the $m\times m$ unit matrix and $P_{{N \over m}}$ 
and $Q_{{N \over m}}$ are the usual building blocks for the twist eating 
configurations:
\begin{equation}
\label{PandQ}
(P)_{ab}=\delta_{a+1,b} \hskip 0.5cm ; \hskip 0.5cm
(Q)_{ab}=\delta_{ab}\ee{\ii{2\pi  m \over N} a} \hskip 0.5cm
,\hskip 0.5cm  a,b=1,\ldots N/m
\end{equation}
with periodicity in the the indices $a $ and $b$, namely   
$a={N \over m}+1$ means  $a=1$.
 It is clear from the context that we have to restrict the values of $N$
so that $N/m$ is an integer.
With these vacuum configurations the trace of open lines is proportional
to $\trace (Q_{{N \over m}})^t$, where $t$ is the difference between the
number of $V$'s and $\Vdag$'s in the trace, namely the difference 
between the time coordinate of the initial and the final point of the 
path in original unreduced lattice.
It is elementary to see from eq. (\ref{PandQ}) that 
\begin{equation}
\label{traceqt}
\trace (Q_{{N \over m}})^t = 0  \hskip 10pt \mbox{unless} \hskip 10pt t=k 
{N \over m}
\end{equation}
for integer $k$. In the unreduced lattice the closed loops correspond to 
$t = k L$, so the comparison of the two equation determines $m$ :
\begin{equation}
\label{emme}
m = {N \over L}~~.
\end{equation}
With the above replacement eq. (\ref{ttekaction}) is, in the large $N$ limit,
an exact dimensional reduction of (\ref{finitet}) on an infinite 
$d$-dimensional lattice.
In order to obtain the matrix model (\ref{model}) that we intend to study,
two more steps are required. First, we assume that an $\vec x$-independent
master field dominates the functional integral so that the field variables
in (\ref{ttekaction}) can be replaced by constants; 
second, we neglect, according
to the previous discussion, the contributions of the space-like plaquettes
by setting $\beta_s = 0$; 
correspondingly, from now on we write simply $\beta$ for $\beta_t$.
For $d=3$ the final result coincides with the one obtained directly from
the hot twisted Eguchi-Kawai model, and the same can be shown to happen 
for any even space-time dimension. 
The derivation in the present section extends the validity
of the reduction to odd space-time dimensions.

The vacuum configuration given in (\ref{ttekvacuum}) and
(\ref{PandQ}) consists
for  $V(\vec x)$ of $L$ bunches of ${N \over L}$ degenerate eigenvalues 
at the $L^{\rm th}$ roots of $1$.
This means that the open Polyakov loop $\left[V(\vec x)\right]^L$
coincides in this extreme weak coupling limit with the unit
matrix\footnote{More generically, we could in (\ref{ttekvacuum})
replace ${\bf 1}_m \equiv {\bf 1}_{N\over L}$ with an element of
${\bf Z}_{N\over L}$. Correspondingly, $V(\vec x)^L$ would be an
element of ${\bf Z}_N$, which is indeed the generic situation in the
broken phase.}.
In the next section the quadratic  quantum fluctuations around this 
vacuum are 
considered and their contribution is  calculated, resulting into a 
broadening
of the eigenvalue distribution within each bunch and eventually, as $\beta_t$
decreases, into a phase transition towards a uniform eigenvalue distribution.
\iffigs
\begin{figure}[t]
\epsfxsize 12cm
\begin{center}
\null\hskip 1pt\epsffile{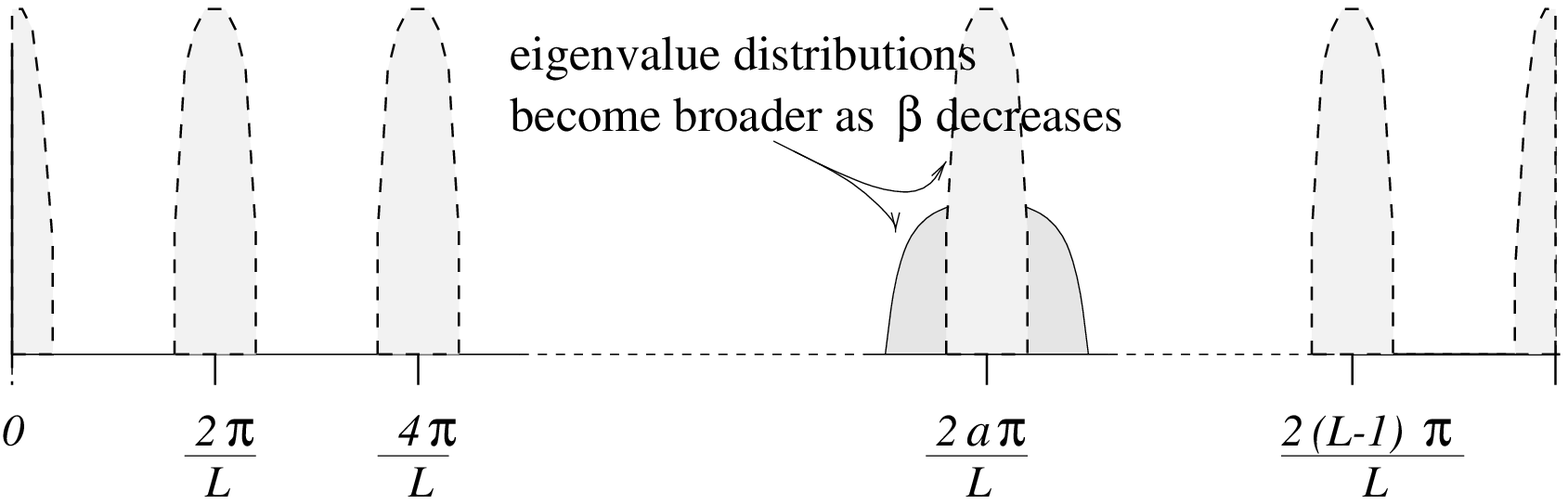}
\vskip 3pt\noindent
\end{center}
\vskip -1.1cm
\mycaptionl{Fig. 1}{The weak coupling solution in the large-${N\over L}$
consists of $L$ 
distributions, centered at the multiples of ${2\pi\over L}$.
The width of the distributions grows as
$\beta$ decreases, until finally an instability is attained at a certain 
critical value $\beta_{\rm cr}$.}
\label{solut}
\end{figure}
\fi 
\section{Derivation of the effective model}
\label{quadmodel}
Let us consider the partition function (\ref{model}) of the twisted 
one-plaquette model.
The integrations over the unitary matrices $U_\mu$ are all independent 
and the partition function can then be rewritten  as
\begin{eqnarray}
\label{wctekpartfunc}
Z &=& \int \hskip -2pt DV \left[\int \hskip -2pt DU\,\, 
\ee{\beta N \real \ee{\ii{2\pi\over L}}
\trace (V U \Vdag \Udag)}\right]^d\nonumber\\
&=& \int\bigl(\prod_{i=1}^N \dop \theta_i\bigr)
\hskip 3pt {\cal J}^2(\theta)
\left[\int\hskip -2pt DU\,\,\ee{\beta N \sum_{i,j} |U_{ij}|^2 \cos (\theta_i -
\theta_j + {2\pi\over L})}\right]^d~~,
\end{eqnarray}
where in the last line we have gauge-rotated the matrix $U$ to
diagonalize $V$: 
\begin{equation}
\label{nn1}
V\rightarrow \diag (\ee{\ii\theta_0},\ldots,\ee{\ii\theta_{N-1}})~~.
\end{equation}
$\cJ(\theta)$ is the
Haar measure of SU(N) expressed in terms of integration over the
eigenvalues $\ee{\ii\theta_i}$ and it is given by
\begin{equation}
\label{cv}
{\cal J}^2(\theta) = \prod_{i<j}4\sin^2 ({\theta_i -\theta_j\over 2})~~.
\end{equation}
We want now to solve the model (\ref{wctekpartfunc}) in the weak
coupling limit, by expanding around the vacuum configuration 
in which\footnote{Here again we assume  $N \over L$ 
to be an integer. This assumption
is justified, as in the Twisted Eguchi-Kawai model, by the belief that
the large $N$ and the continuum limit (large $L$) are smooth. 
It would nevertheless be of some interest  to study the case of  
$N = m L + r$ with $m$ and $r$ integers and $0<r<L$. 
Although sub-leading in the large $N$ expansion, the dynamics of the
$r$ eigenvalues that cannot accommodate into $L$ identical bunches
might reveal some interesting features, and it is likely to depend
on the existence of common divisors of $L$ and $r$. }
\begin{equation}
\label{conto1}
V=Q_L \otimes {\bf 1}_{N/L}
\hskip 0.8cm ; \hskip 0.8cm 
{\cal P}={1\over N}\trace V^L=1 ~~.
\end{equation}
In this vacuum the eigenvalues of $V$ are organised in $L$ bunches,
each composed of ${N \over L}$ identical values $\ee{2\pi\ii a\over L}$,
$a =1,\ldots L$.
Accordingly we write the eigenvalues of the matrix $V$ in the weak
coupling phase as
\begin{equation}
\label{conto3}
\theta_i \equiv  
\theta_{a,\alpha} =  {2\pi a\over L} + \phi_a + \varphi_{a,\alpha}~~,
\end{equation}
with $a =1,\ldots L$ , $\alpha=1,\ldots {N\over L}$ and with 
$\varphi_{a,\alpha}$ constrained by $\sum_\alpha \varphi_{a,\alpha} = 0$.
We have separated here the degrees of freedom $\phi_a$ corresponding 
to the fluctuations of the centres of the bunches from the fluctuations
$\varphi_{a,\alpha}$ within each bunch.  

We will show that in the large $N/L$ limit the model is solved, for $\beta$
bigger than a calculable critical value $\beta_{\rm cr}$, by an
eigenvalue distribution for $V$ consisting of L separated distributions, all
centered at a multiple of ${2\pi\over L}$. In the large $L$ limit 
we will show that the $L$ bunches of eigenvalues all have the same Wigner 
semicircular distribution.
\subsection{Integration over the space-like links}
The first step in order to solve the model is to perform explicitly the 
integration over the unitary matrix
$U$ in the expression (\ref{wctekpartfunc}) of the partition function. This
will lead us to an ``effective'' action for the eigenvalues of the matrix 
$V$, which  will be studied by means of  matrix models techniques.

Let us consider the slightly more general integral
\begin{equation}
\label{int}
I(\theta,\psi) = \int\hskip -2pt DU\,\, 
\ee{ \beta N \sum_{ij} \cos (\theta_i - \psi_j) |U_{ij}|^2}~~,
\end{equation}
which reduces to the one of (\ref{wctekpartfunc}) for 
$\psi_j = \theta_j - {2 \pi \over L}$.

In spite of its analogy with  the Itzykson-Zuber
integral, the integration in (\ref{int}) cannot be done 
exactly (except for $N=2$) as the Duistermann-Heckmann theorem 
does not apply in this case. However  in the large $N$ limit 
it is possible to apply to it the saddle-point
method. In this way Kogan et. al \cite{Kogan}, by taking into account, 
like in the Itzykson-Zuber
integral, all extrema -- both maxima and minima -- of the exponent, 
obtained the following asymptotic formula:
\begin{equation}
\label{kogan}
\int\hskip -1pt DU \,\ee{\sum_{ij} A_{ij} |U_{ij}|^2}
= \sum_P {\exp\{\sum_k A_{k\, P(k)}\}\over
\prod_{i<j} [A_{i\, P(i)} + A_{j\, P(j)} - A_{i\, P(j)}
-A_{j\, P(i)}]}\hskip 3pt (1 + O({1\over A}))~~,
\end{equation}
$P$ being a permutation of the indices $\{i\}$. By replacing in 
(\ref{kogan}) $A_{ij}$ with $\beta N \cos (\theta_i -\psi_j)$ and by 
using the trigonometrical identity
\begin{eqnarray}
\label{trig}
&\cos (\theta_i - \psi_{P(i)}) + \cos (\theta_j - \psi_{P(j)}) - \cos 
(\theta_i - \psi_{P(j)}) - \cos (\theta_j - \psi_{P(i)})\nonumber \\ 
&= 4 \sin ({\theta_i - \theta_j \over 2}) 
\sin({\psi_{P(i)}-\psi_{P(j)} \over 2})
\cos ({\theta_i +\theta_j-\psi_{P(i)}-\psi_{P(j)} \over 2})
\end{eqnarray}
we would obtain the following expression for the large $N$ limit of
$I(\theta,\psi)$:
\begin{equation}
\label{largenint}
I(\theta,\psi) =C  \sum_P (-1)^P {\cal J}^{-1}(\theta) {\cal J}^{-1}(\psi) 
{\exp\{ N \beta \sum_k 
\cos(\theta_k - \psi_{P(k)})\} \over \prod_{i,j} \cos \bigl({\theta_i - 
\psi_{P(i)} + \theta_j - \psi_{P(j)} \over 2 }\bigr)} (1 + O({1 \over N}))~,
\end{equation}
where  $C=(\beta N)^{N(N-1)\over 2}$ is an irrelevant overall factor,
which will be discarded in what follows.

In applying the saddle point method to the integral at the l.h.s. of
eq. (\ref{kogan}) one should be aware that the gaussian integration
around the extrema would diverge as a maximum turns into a minimum of 
the integrand. These divergences are signalled by the zeros at the denominator
of the r.h.s of eq. (\ref{largenint}). The zeros of ${\cal J}(\theta)$ and
${\cal J}(\psi)$ however are cancelled by the zeros at the numerator when the
sum over permutations is performed, in analogy to what happens in the 
Itzykson-Zuber integral. Instead the zeros of $\cos \bigl({\theta_i -
\psi_{P(i)} + \theta_j - \psi_{P(j)} \over 2 }\bigr)$ are true singularities
of the r.h.s of (\ref{largenint}). These singularities are not present in the
original integral and they denote that the equation (\ref{largenint}) cannot 
be applied in such circumstances unless one restricts the sum to a suitable
subset of permutations.

Let us go back to our original integral of eq. (\ref{wctekpartfunc}), which
is obtained from (\ref{largenint}) with the substitution $\psi_i \rightarrow
\theta_i - {2 \pi \over L}$. 
The vacuum configuration of such a model was discussed in the previous section
and it is given by eq. (\ref{ttekvacuum}), with $N/m = L$.

To consider just  quantum fluctuations around this vacuum 
is equivalent to restrict the sum over permutations at the r.h.s. of
(\ref{largenint}) to the permutations that map each bunch of eigenvalues
into the next bunch, namely:
\begin{equation}
\label{joe4}
P:\hskip 0.2cm \theta_{a,\alpha} \rightarrow \theta_{a+1,P_a(\alpha)}~~.
\end{equation}
where $P_a(\alpha)$ corresponds to a permutation of the ${N \over L}$ 
eigenvalues of the bunch at ${ 2 \pi a \over L}$.
The other permutations are exponentially depressed compared to the ones
above; in fact for each eigenvalue mapped from the bunch 
$a$ to the bunch $a+s+1$
the exponential in (\ref{largenint}) is depressed by a factor
\begin{equation}
\label{conto5}
\exp \left(\beta N(1-\cos {2\pi s\over L})\right) 
\sim \exp\left(-\beta N {2 \pi^2 s^2\over L^2}\right)~~.
\end{equation}
In the next section we shall precisely examine the 
quantum fluctuations around the classical vacuum given in (\ref{ttekvacuum}), 
and determine for what value
of $\beta$ such fluctuations make the vacuum unstable. In other words we shall
neglect all extrema of the action other than the ones given in eq. 
(\ref{joe4}), and also keep the center of the bunches fixed at their classical
value, namely we shall fix $\phi_a =0$ through all the calculation.
Of course the question arises whether a condensation of the other extrema, 
or an instability in the dynamics of the bunches as a whole may cause a 
transition {\it before} the vacuum becomes unstable due to the quantum 
fluctuations.
In the next section we shall give strong arguments, based on the comparison of
the present model with QCD in two dimension on a cylinder, 
that at least for large $L$ the value of $\beta$ for which the vacuum 
becomes unstable in very close, if not coincident, with the one where the 
transition takes place.
Finally, let us remark that when the sum over permutations in 
eq. (\ref{largenint}) is restricted to the ones of (\ref{joe4}) 
the result of the saddle point method is singularity free, as the argument 
of the cosine at the denominator is for those permutation close to zero.
\subsection{The effective quadratic model}
With the replacement $\psi_i \rightarrow \theta_i - {2 \pi \over L}$ the r.h.s.
of eq. (\ref{largenint}) becomes
\begin{equation}
\label{lnexp}
 \sum_P (-1)^P {\cal J}^{-2}(\theta) 
{\exp\{ N \beta \sum_k 
\cos(\theta_k - \theta_{P(k)} + {2 \pi \over L})\} \over \prod_{i,j} 
\cos \bigl({\theta_i - 
\theta_{P(i)} + \theta_j - \theta_{P(j)} \over 2 }\bigr)}~~,
\end{equation}
where the sum is now restricted to the permutations given in eq. (\ref{joe4}).
We replace $\theta_i$ with its classical value plus fluctuations, according
to eq. (\ref{conto3}), and expand each term up to quadratic order in the
fluctuations.
It is easy to see that upon this expansion the fluctuations $\phi_a$ 
(the ``centres'') and $\varphi_{a,\alpha}$ decouple completely. As already
discussed, we consider the solution where all the $\phi_a$ vanish. In the 
remaining of this section we keep therefore track only of the terms containing
 $\varphi_{a,\alpha}$ and we  derive a quadratic 
effective model for them.

Let us consider first the factor  ${\cal J}^2(\theta)$ appearing both in the
integration volume and in (\ref{lnexp}). Under
the substitution (\ref{conto3}) it becomes
\begin{eqnarray}
\label{nuova2}
&&{\cal J}^2(\theta) =\prod_{ij} 4 \sin^2 {\theta_i - \theta_j\over 2} 
\rightarrow \prod_{a,\alpha < b,\beta}
4\sin^2 \biggl[ {\pi(a-b)\over L} + {\varphi_{a,\alpha} - \varphi_{b,\beta} 
\over 2}\biggr]
\nonumber\\
&& = \Bigl[\prod_a \Delta^2(\varphi_a)\Bigr] \,\times\, 
\exp\Biggl\{ \sum_{a\not= b}
\sum_{\alpha\beta} \log \Bigl(2\sin \biggl[{\pi(a-b)\over L} + 
{\varphi_{a,\alpha}- \varphi_{b,\beta} \over 2} \biggr]\Bigl) 
\nonumber\\
&&\hskip 3.8cm + \sum_a \sum_{\alpha\not=\beta} 
\log {\sin {\varphi_{a,\alpha} - \varphi_{b,\beta} \over 2} \over 
{\varphi_{a,\alpha} - \varphi_{b,\beta} \over 2} }\Biggr\}~~,
\end{eqnarray}
where $\Delta^2(\varphi_a) = \prod_{\alpha < \beta} (\varphi_{a,\alpha} -
\varphi_{a,\beta})^2$ is the usual Cauchy-Vandermonde determinant.

The quadratic expansion of eq. (\ref{nuova2}) is
\begin{equation}
\label{nuova3}
\Bigl[\prod_a \Delta^2(\varphi_a)\Bigr] \,\times\, 
\exp\Bigl\{-{1\over 4} {N\over L} \Bigl(\sum_{l=1}^{L-1} 
{1\over \sin^2 {\pi l\over L}}
+ {1\over 12} \Bigr)
\sum_{a,\alpha} \varphi_{a,\alpha}^2 + O(\varphi^4)\Bigr\}~~.
\end{equation}
The finite sum appearing above can be exactly calculated  
(see App. \ref{appea}) and it is given by
\begin{equation}
\label{conto11}
\sum_{l=1}^{L-1} {1\over \sin^2 {\pi l \over L}} = {L^2-1\over 3}~~.
\end{equation}     
The exponential and the denominator in (\ref{lnexp}) become respectively
\begin{eqnarray}
\label{conto6}
&&\exp\Bigl\{\sum_i N \beta  \sum_k 
\cos(\theta_k - \theta_{P(k)} + {2 \pi \over L})\}\Bigr\} \rightarrow
\exp\Bigl\{\beta N\sum_{a,\alpha} \cos (\varphi_{a,\alpha}-
\varphi_{a+1,P_a(\alpha)})\Bigr\} \nonumber \\ &&= 
\exp\Bigl\{ - \beta N (\sum_{a,\alpha} \varphi_{a,\alpha}^2 
- \sum_{a,\alpha} \varphi_{a,\alpha} \varphi_{a+1,P_a(\alpha)} + O(\varphi^4))
\Bigl\} 
\end{eqnarray}
and
\begin{eqnarray}
\label{wctekdenominator}
&&\prod_{i<j} \Bigl(
\cos {\theta_i - \theta_{P(i)} + \theta_j - \theta_{P(j)}
\over 2} \Bigr)^{-1}\nonumber\\
&&\rightarrow \exp\Bigl\{ -\sum_{a,\alpha < b,\beta}  \log \cos \bigl[
{\varphi_{a,\alpha} - \varphi_{a+1,P_a(\alpha)} + \varphi_{b,\beta} - 
\varphi_{b+1,P_b(\beta)} \over 2}   \Bigr]  \Bigr\}
\nonumber\\ &&=\exp\Bigl\{ {N\over 4} \Bigl(\sum_{a,\alpha} 
\varphi_{a,\alpha}^2 - 
\sum_{a,\alpha}\varphi_{a,\alpha}\varphi_{a+1,P_a(\alpha)} + O(\varphi^4)\Bigr)
\Bigr]\Bigr\}~~.
\end{eqnarray}
Putting all these results together, we can now write
down the explicit expression of the action (\ref{wctekpartfunc}) 
expanded  up to quadratic terms
in the fluctuations $\varphi$:
\begin{eqnarray}
\label{nuova33}
&& \int \Bigl[ \prod_{a\alpha} {\rm d}\varphi_{a,\alpha} \Bigr] 
\Bigl[\prod_{a}
\Delta^2(\varphi_a)\Bigr]^{(1-d)} \,\,\exp\Bigl\{ -{N\over L} \Bigl[ 
(\beta L- {L\over 4}) d - {d-1\over 12}L^2 \Bigr]
\sum_{a,\alpha} \varphi_{a,\alpha}^2\Bigr\}
\nonumber\\
&& \hskip 1.2cm\times \Biggl( \sum_{\{P_a\}} (-1)^{\sigma(P_a)} \, 
\exp\Bigl\{ {N\over L} (\beta L - {L\over 4}) \sum_{a,\alpha} 
\varphi_{a,\alpha} \varphi_{a+1,P_a(\alpha)} \Bigr\} \,\Biggr)^d~~.
\end{eqnarray}
The solution of this model is the subject of the next section.
\section{Solution of the model}
\label{solution}
\subsection{${\bf Z}_L$ invariant solution}
\label{sollargel}
In solving the model (\ref{nuova33}) for an arbitrary value of $L$, 
we assume that the master field is translational invariant, 
namely invariant under the ${\bf Z}_L$ symmetry of the 
vacuum  (\ref{conto1}). With this assumption
all bunches of eigenvalues have the same distribution and we can set:
\begin{equation}
\label{conto21}
\varphi_{a,\alpha} \rightarrow {\varphi_\alpha\over L}~~,
\end{equation}
where $\varphi_\alpha$ is now of order $1$ in the large $L$ limit (recall that
the fluctuations $\varphi_{a,\alpha}$ 
are  of order $1\over L$). The invariance under translations is 
at this stage an assumption, although a very reasonable one; it will
be checked {\it a posteriori} in section \ref{Lgeneric}  for an 
arbitrary value of $L$.
 
With the position (\ref{conto21}) the partition function (\ref{nuova33})
reduces to the product of the partition functions of $L$ identical models:
\begin{eqnarray}
\label{conto22}
Z =(Z_{KM})^L & = & \Biggl[\int\prod_{\alpha=1}^n {\rm d}\varphi_\alpha\, 
\Bigl[\Delta^2(\varphi)\Bigr]^{1-d}
\,\,\exp\Bigl\{-n \Bigl[\beta_H d - {d-1\over 12}  
\Bigr] \sum_\alpha(\varphi_\alpha)^2
\Bigr\}\nonumber\\
&&\times\Biggl(\sum_P (-1)^{\sigma(P)} \exp\Bigl\{
n \beta_H \sum_\alpha 
\varphi_\alpha\varphi_{P_a(\alpha)}\Bigr\}\Biggr)^d \Biggr]^L \,\, ,
\end{eqnarray}
where we have defined
\begin{equation}
\label{conto23}
\beta_H = {1\over L}(\beta -{1\over 4})~~,
\end{equation}
and $n = N/L$.
The partition function of each individual model
in the above equation coincides with the one of  a 
$d$-dimensional Kazakov-Migdal model \cite{km} with
quadratic potential. In fact from the Kazakov-Migdal partition function
\begin{eqnarray}
\label{conto24}
&& \int \prod_{\mu=1}^d \prod_{\vec x} DU_\mu(\vec x) \int \prod_{\vec x}
D\Phi(\vec x)\,\, \exp\Bigl\{-n\beta_H \trace \sum_{\vec x}\Bigl[
{m^2\over 2} \Phi^2(\vec x) 
\nonumber\\ 
&&\hskip 4cm
- \sum_{\mu=1}^d \Phi(\vec x) U_\mu(\vec x) \Phi(
\vec x + \mu) U^\dagger_\mu(\vec x)\Bigr]\Bigr\}
\end{eqnarray}
by integrating over the unitary matrices $U_\mu(\vec x)$ one obtains in the 
mean-field approximation  ($\Phi(\vec x)$ independent of $\vec x$) the
expression $Z_{KM}$ in (\ref{conto22}),
 where $\{\varphi_\alpha\}$ are the eigenvalues 
of the hermitean matrix $\Phi$, and the mass $m$ is:
\begin{equation}
\label{conto25}
m^2 = 2 d -{d-1\over 6 \beta_H}~~.
\end{equation}
The model (\ref{conto24}) was solved in the large-$n$ limit by 
Gross \cite{gross},
who showed that the  eigenvalue distribution $\rho(\varphi)$ is 
semicircular:
\begin{equation}
\label{semicirc}
\begin{array}{ll}
{\displaystyle\rho(\varphi) = {2\over \pi r^2} \sqrt{r^2 - \varphi^2}}
\hskip 0.5cm & \mbox{for}\,\, |\varphi | < r\, , \cr
\null & \null \cr
{\displaystyle\rho(\varphi)} = 0 & \mbox{for}\,\, r < |\varphi | < \pi
\end{array}
\end{equation}
 with radius given by
\begin{equation}
\label{conto26}
r^2 = {4(2d-1)\over \beta_H \biggl(m^2 (d-1) + d\sqrt{m^4 - 4(2d-1)}
\biggr)}~~.
\end{equation}
The solution becomes unstable at the value of 
$\beta_H$ for which the radius become complex, namely 
\begin{equation}
\label{conto27}
\beta_H = {d-1\over 12 (d - \sqrt{2d-1})}~~.
\end{equation}
The corresponding value of $\beta$ (recall the definition (\ref{conto23}))
represents a lower bound for the true critical coupling $\beta_{\rm cr}$ 
at which a phase transition  takes place.

It should be noticed  however that a different upper limit may be found by 
looking at the value
of $\beta_H$ for which the radius of the eigenvalue distribution of
$\varphi_{\alpha}$ becomes
$\pi$,  corresponding to a radius ${\pi \over L}$ for each bunch
of eigenvalues. A phase transition occurs at this point as the bunches merge 
together and the semicircular distribution is not anymore a solution.

A better understanding of the 
phase transition can be obtained by noticing that for large $L$, 
namely in the continuum limit, the Wilson action becomes equivalent to the 
heat-kernel action~\cite{drouffe,menon}. To be more precise, if we 
consider the action (\ref{finitet})
with $\beta_s=0$, in the large $L$ and hence large $\beta_t$ limit we have
\begin{eqnarray}
&&\exp\left(N \beta \sum_{\vec{x}} \sum_{t=1}^{L} 
\sum_{\mu =1}^d {\rm Re} {\rm Tr}  U_{\mu}(\vec{x},t) 
V(\vec{x}+\hat{\mu},t) U^\dagger_{\mu}(\vec{x},t+1)
V^\dagger(\vec{x},t)\right)\,  
\stackrel{\beta \rightarrow \infty}{\longrightarrow} \nonumber \\ 
&&\prod_{\vec{x},t,\mu} \left\{ \sum_r d_r \chi_r 
(U_{\mu}(\vec{x},t) V(\vec{x}+\hat{\mu},t) U^\dagger_{\mu}(\vec{x},t+1)V^\dagger
(\vec{x},t)  ) \exp \left(- {C_r \over 2 N \beta}\right) \right\}~,
\label{hamlim}
\end{eqnarray}
where $r$ labels the irreducible representations of $SU(N)$, $d_r$ their
dimensions, $C_r$ their quadratic Casimir and $\chi_r(U)$ 
the corresponding character of $U$. Consistently with previous notations
we  denote $\beta_t$ by $\beta$ in (\ref{hamlim}) and following equations.
The heat-kernel action at the r.h.s. of (\ref{hamlim}) was studied in detail
in ref.s ~\cite{largeN} and ~\cite{zarembo}, which papers we 
refer the reader to for further details. 
The integration over the space-like links can be done 
exactly by using well known properties of characters, leading to the 
following action for the Polyakov loop: 
\begin{equation}
\exp (S_{\rm Pol}) = \prod_{\vec{x},\mu}
\sum_r \chi_r({\cal P}(\vec{x}+\hat{\mu})) \chi_r({\cal P}^\dagger(\vec{x})) 
\exp \left(- {C_r L\over 2 N \beta}\right)~~,  
\label{zeta0}
\end{equation}
where ${\cal P}(\vec{x})=\prod_{t=1}^L V(\vec{x},t)$ 
is the untraced Polyakov loop.
By assuming as usual that the master field is invariant under translations,
it was shown in ~\cite{largeN} that from (\ref{zeta0}) one obtains 
the following partition function:
\begin{eqnarray}
\label{conto28}
Z & = & \int\prod_{i=1}^N {\rm d}\hat\theta_i\,
\Bigl[\Delta^2(\hat\theta)\Bigr]^{1-d}\,\,
\exp\Bigl\{-N \biggl[{\beta \over L} d - {d-1\over 12}\biggr]
\sum_i(\hat\theta_i)^2 \Bigr\} \nonumber\\ 
& &\times\Biggl[\sum_p (-1)^{\sigma(P)} 
\,\exp\Bigl\{N {\beta \over L}\, 
\hat\theta_i\hat\theta_{Pi}\Bigr\} \Biggl]^d~~.
\end{eqnarray}
The integration variables $\hat\theta_i$ in eq. (\ref{conto28}) 
are the invariant angles of the Polyakov loop, 
and higher order terms in $\hat\theta_i$ in the exponent have 
been neglected. The partition function given in (\ref{conto28}) is again a 
Kazakov-Migdal model with a quadratic potential, and it admits the same 
semicircular distribution (\ref{semicirc})  as the partition function
(\ref{conto22}) obtained from our one plaquette model, but with $\beta_H$
replaced by $\beta/L$.
This is in complete agreement with the leading order in $L$ of the 
rescaling law given by eq. (\ref{conto23}).
The only approximation involved in deriving eq. (\ref{conto28}) 
is the truncation of the power expansion in $\hat\theta_i$ at the 
exponent to the quadratic terms.
In conclusion, the twisted one plaquette model and the heat-kernel model
have in the large $L$ limit the same solution for the eigenvalue distribution
of the Polyakov loop, and the same critical value of the coupling constant.
This confirms that the local extrema which were neglected in the 
saddle point evaluation of (\ref{int}) are indeed irrelevant, at least 
in the large $L$ limit and up to the critical value of the coupling constant.

The relation between our one plaquette model and the 
heat-kernel model discussed in~\cite{largeN} also provides a new 
and {\it a priori} independent way of estimating 
the critical coupling. In fact it was shown in~\cite{largeN}, 
and it can be recognised in (\ref{zeta0})\footnote{The expression 
under the product sign in (\ref{zeta0})
is the partition function for QCD2 on on a cylinder with boundary conditions
given by the Polyakov loops in $\vec x$ and ${\vec x}+{\hat \mu}$.}, 
that in the heat-kernel model the building
block  of the effective action for the Polyakov loop is given by the action of 
QCD2 on a cylinder of area inversely proportional to $\beta/L$.
So the critical value of $\beta_H\sim \beta/L$ should correspond to the one
where the  Douglas-Kazakov phase transition~\cite{dk} occurs.
The point of the Douglas-Kazakov phase transition is known exactly as a 
function of the area of the cylinder and of the radius of the semicircular 
distributions at its  boundaries. The radius of the distribution is obtained 
from the solution of (\ref{conto28}) and it is given, as before, 
by (\ref{conto26}). The critical coupling is determined now by the radius where
the Douglas-Kazakov phase transition occurs (see eqs. (33) and (42) in 
~\cite{largeN}) rather than by looking at the point where 
the radius becomes complex.
The results of  Table I  show that all the above  methods  give, at
least for low $d$,  values for the critical  
$\beta_H$ which are very close to each other, and that the ones obtained
from the analysis of  the Douglas-Kazakov phase transition give consistently 
a better (higher) lower limit.
\begin{table}
\mytcaptionl{Table I}{Critical value of $\beta_H$ calculated in 
different ways: (a) limiting value for real radius, (b) limiting value 
for $ r < \pi$ and (c) value corresponding to the DK phase transition in 
QCD2 \cite{largeN}. For $d>3$ the solution becomes unstable before the radius 
reaches the value $\pi$.}
\label{tab1}
\vskip -0.3cm
\begin{center}
\begin{tabular}{c c c c c c}
\hline\hline
 &  (a) & (b) & (c)  \\
\hline
$d=2$   &0.311  &0.314  &0.321 \\
$d=3$   &0.218  &0.218  &0.226 \\
$d=4$   &0.184  & *     &0.192 \\
$d=5$   &0.167  & *     &0.173 \\  
\hline\hline
\end{tabular}
\end{center}
\end{table}

It is well known that the Douglas-Kazakov phase transition 
is due to the condensation of ``instanton"
contributions, namely to configurations where the initial and final value of 
one or more eigenvalues of ${\cal P}$ differ of more than $2 \pi$. 
We argue that there is a precise correspondence between the instantons in QCD2 
on a cylinder and the classical solutions of the one plaquette model (extrema 
of the integral $I(\theta, \theta - {2 \pi \over L})$) which were
neglected in the saddle point  calculation of the previous section.
These correspond to permutations which are not of the type of eq. (\ref{joe4}), 
because one or more eigenvalues of $V$ are not mapped from a bunch 
$a$ to $a + 1$.
The argument goes as follows: in the large $L$ limit the eigenvalue distribution
of our solution can be represented on an infinite line as a 
sequence of bunches, each made of $n=N/L$ eigenvalues\footnote{We refer 
here to the eigenvalues normalised as $\varphi_{\alpha}$  in (\ref{conto21}).},
at intervals of $2 \pi$. 
Above the phase transition  different bunches do not communicate
with each other, except for the canonical shift of $1$ due to the twist, 
and the partition function is just the one corresponding to a single 
bunch raised to  the power $L$ (see eq. (\ref{conto22}) ).
The picture is quite similar for the eigenvalues of the Polyakov loop in 
the heat-kernel model. Their distribution can be represented on an 
infinite line as a periodic distribution of period $2 \pi$, that before 
the phase transition consists of separate bunches of $N$ eigenvalues. 
Let us compare now the contribution of  classical solutions of the 
instanton type in the two models.
Instanton contributions to the partition function of QCD2 on
a cylinder can be calculated  from the following  equation 
(see eq. (8) in ~\cite{dkcyl})
\begin{eqnarray}
\label{instan}
&  & \sum_r  \chi_r(-\hat\theta) \chi_r(\hat\varphi)  
\exp \left( - { t C_r \over 2} \right) \nonumber \\
&\propto& \sum_P {t^{(1-N)/2} \over {\cal J}(\hat\theta) {\cal J}(\hat\varphi)}
\sum_{\{s_i\}} \exp \left[- {1 \over 2 t} \sum_{i=1}^N 
(\hat\varphi_i - \hat\theta_{P(i)} + 2 \pi s_i)^2 \right]~~. 
\end{eqnarray}
where $t$ can be identified  with ${L \over N \beta}$, by comparing eq. 
(\ref{instan}) with (\ref{zeta0}).
The integers $s_i$ correspond to the winding numbers of the eigenvalues, 
so the contribution of an instanton with winding number $s$ is proportional 
to $\exp (- 2 \pi^2 \beta N s^2 / L)$.

Let us compare this result with  the non perturbative 
contribution  (\ref{conto5}) corresponding to an eigenvalue being mapped by a 
permutation $P$ from a bunch $a$ to the bunch $a+s+1$
in the twisted one plaquette model.
The r.h.s. of (\ref{conto5})  coincide with the instanton 
contribution calculated above, if one takes into account that the
number of eigenvalues is here $n=N/L$ instead of $N$.

We can  conclude that the mechanism underlying the phase transition is the same
in QCD2 on a cylinder and in the twisted one plaquette model.   In 
particular in the latter model only 
fluctuations around the twist eating configurations need to be considered 
for $\beta$ larger than its critical value; when the critical value is
attained a condensation of local extrema of the action takes place, that 
is responsible for the phase transition studied in this section.
\subsection{Quantum rescaling}
\label{qresc}
Let us consider again the rescaling law given by eq. (\ref{conto23}),
and write it as:
\begin{equation}
\label{conto30bis}
\beta(L) = L \beta_H + {1 \over 4}~~,
\end{equation}
where in the l.h.s. we replaced $\beta$ with $\beta(L)$ to remark the fact
that the coupling refers to a model with $L$ time-like links.
On the other hand we have $L= \rho n_t$ (see footnote in
section 2.2), where $\rho$ is the asymmetry parameter and $n_t$ the number of
links in the time direction in the equivalent symmetric lattice.
By writing eq. (\ref{conto30bis}) with $L=\rho n_t$  and with $L= n_t$, and 
by eliminating $\beta_H $ from the two equations, we find
\begin{equation}
\label{conto30ter}
\beta(\rho n_t) = \rho \biggl(\beta(n_t) - {1 \over 4}\biggr) + {1 \over 4}~~.
\end{equation}
This equation gives the riscaling of the coupling constant induced by
varying the asymmetry parameter $\rho$, and it should be compared with
the first of eqs. (\ref{asymmetry}) where $\beta_{\rm symm} \rightarrow 
\beta(n_t)$ and $\beta_t(\rho) \rightarrow  \beta(\rho n_t)$. 
The function
$c_{\tau}(\rho)$ of eq. (\ref{asymmetry}) was calculated by Karsch in 
\cite{Karsch}.
Its behaviour for large $\rho$
is of the type
\begin{equation}
\label{karasy}
4 \rho\, c_{\tau}(\rho) = \alpha_t^0 \rho + \alpha_t^1 + 
O\left({1 \over \rho}\right)~~,
\end{equation}
where the values of the coefficients $\alpha^0_t$ and $\alpha^1_t$ in the 
large-$N$ limit can be calculated from the eq. 2.25 of \cite{Karsch}:
\begin{equation}
\label{conto35}
\alpha^0_t = - 0.2609 \hskip 1cm ; \hskip 1cm \alpha^1_t = {1\over 4}~~.
\end{equation}
We can extract for comparison the same quantities from  the rescaling 
(\ref{conto30ter}),  following from our effective model. 
We find : 
\begin{equation}
\label{joe7}
\alpha^0_t = -{1\over 4}  \hskip 1cm ; \hskip 1cm \alpha^1_t = {1\over 4}~~.
\end{equation}
The agreement is quite remarkable, considering that Karsch's calculation took
into account the space-like plaquettes, and it confirms a feature that already
emerged in the case of SU(2)~\cite{bcdp}, namely that the corrections to 
$c_{\tau}(\rho)$ due to the space-like plaquettes are relatively small.
The agreement between (\ref{joe7}) and (\ref{conto35}) also represents a 
further check that only the fluctuations around the classical vacuum are
relevant in the deconfined phase.
\subsection{Evidence for unbroken ${\bf Z}_L$ symmetry}
\label{Lgeneric}
In section 4 we solved the model described by the action (\ref{nuova33}) by
assuming that the solution is translational invariant in the one dimensional
space labelled by the index $a$ in $\varphi_{a,\alpha}$. In other words
we assumed that the $Z_L$ symmetry of the vacuum (\ref{conto1}) is not broken 
by the quantum fluctuations. 
Although this is justified in the large $L$ limit, it would be desirable to have
an independent proof valid for all values of $L$.
In this section we shall prove that for any $L$ the partition function
(\ref{nuova33}) becomes singular at the critical value of $\beta$ 
derived in the previous section, namely at the value of $\beta$ for 
which the radius of the large L semicircular solution becomes complex.

To begin with, let us notice that the partition function (\ref{nuova33}) can be 
obtained by integration \cite{izhc} over the unitary matrices $U$ 
of the following model of Kazakov-Migdal type:
\begin{equation}
\label{conto36}
Z = \int \biggl[\prod_{\mu = 1}^d DU^{(a)}_\mu\biggr]\,
\biggl[\prod_{a=1}^L D\Phi_a\biggr] \, \exp\Bigl\{-n \beta_H
 \trace \Bigl[\sum_a {m^2\over 2} \Phi_a^2 
-\sum_{a,\mu} \Phi_a U_\mu^{(a)} \Phi_{a+1} U_\mu^{(a)\,\dagger}
\Bigr] \Bigr\}~,
\end{equation}
where the matrices  $\Phi_a$ are normalised so that their eigenvalues
are  $L \varphi_{a,\alpha}$, and the parameters $\beta_H$ and $m^2$
are given by eqs. (\ref{conto23}) and (\ref{conto25}).

The structure of the model (\ref{conto36}) is illustrated in Fig. 2; 
it looks like a 
KM model defined on a compact one-dimensional lattice of length $L$, but
with  $d$ links joining two neighbouring sites.
\iffigs
\begin{figure}
\epsfxsize 6cm
\begin{center}
\null\hskip 1pt\epsffile{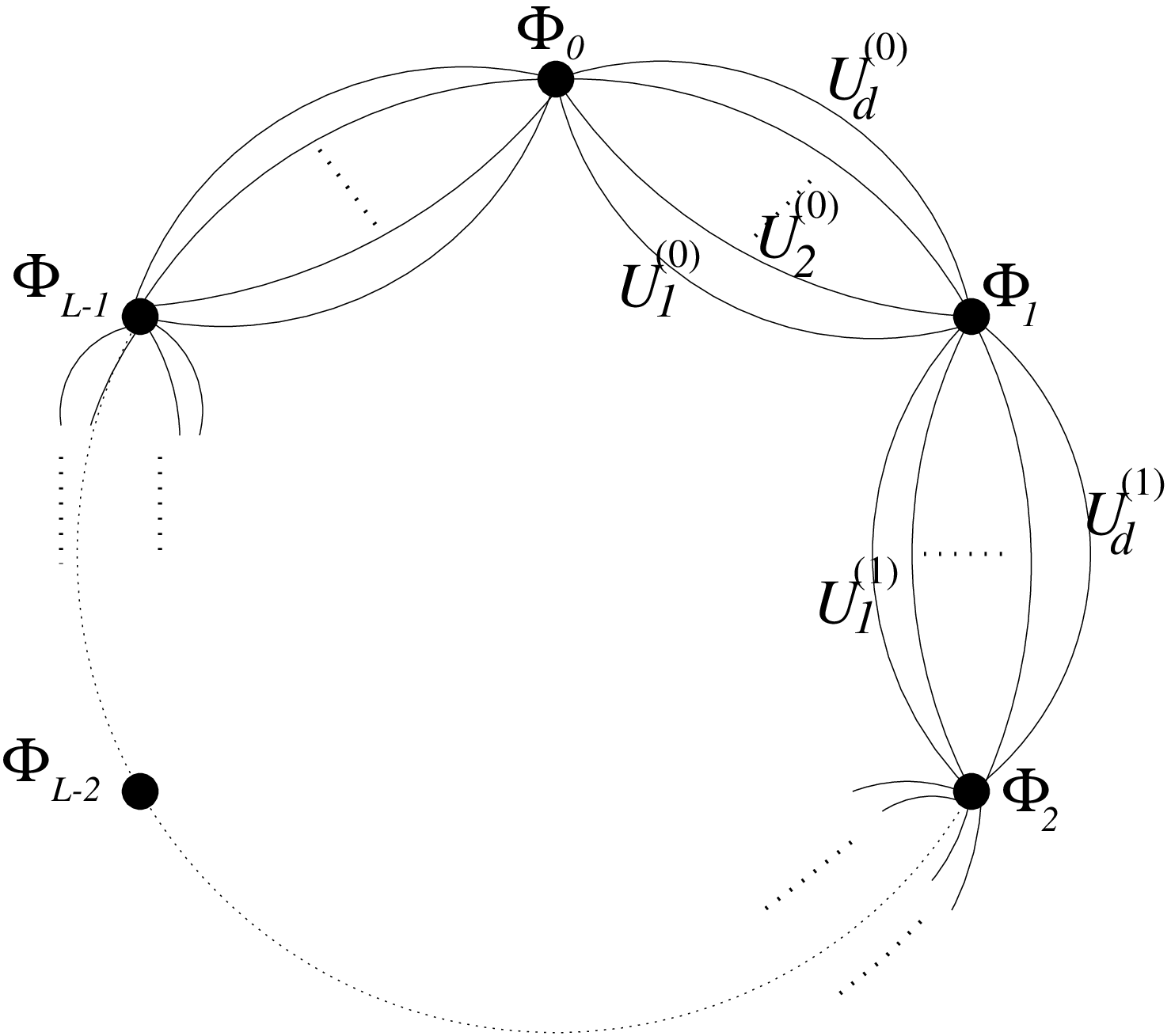}
\end{center}
\vskip -0.6cm
\mycaptionl{Fig. 2}{Structure of the effective KM model corresponding 
to the case of $L$  generic.
Each link $\stackrel{\Phi_a}{\bullet} \stackrel{U_\mu^{(a)}}{\longrightarrow
\longrightarrow}\hskip -3pt
\stackrel{\Phi_{a+1}}{\bullet}$ represents a term $\Phi_a U_\mu^{(a)} 
\Phi_{a+1} U_\mu^{(a)\dagger}$ in the action.}
\label{KMmod}
\end{figure}
\fi

If one integrates eq. (\ref{conto36}) over the hermitean matrices $\Phi_a$ 
rather than over the unitary matrices $U$, one obtains the 
``induced gauge model'' \cite{km}
\begin{equation}
\label{conto37}
Z = \int\biggl[\prod_{a=0}^{L-1}\prod_{\mu = 1}^d DU^{(a)}_\mu\biggr]\,\,
\exp\Bigl\{{1\over 2} \sum_\Gamma {|\trace  U_\Gamma |^2 \over l_\Gamma
(m^2)^{l_\Gamma} } \Bigr\}~~,
\end{equation}
where $\Gamma$ is a closed path on the lattice of Fig. 2 and $l_\Gamma$ 
its length.  
$U_\Gamma$ is the ordered product of the link 
matrices $U_\mu^{(a)}$ on the path  $\Gamma$. 

Consider now the quantity $\rho_a^2={1\over n}\langle\trace \Phi_a^2\rangle$, 
which is a measure of the width of the $a$'th bunch and is proportional, 
in case of a semicircular distribution, to the squared radius $r^2$.
It is easy to show, by using eq. (\ref{conto36}), that
\begin{equation}
\label{conto38}
 \rho_a^2 = {1\over n}\langle\trace \Phi_a^2 \rangle  
={1\over \beta_H}\langle\, \sum_\Gamma
{|\trace U_\Gamma|^2\over n^2 (m^2)^{l_\Gamma + 1}}~~\rangle~~,
\end{equation}
where the sum is over all paths $\Gamma$ beginning and ending in $a$ and
the last  expectation value   is taken 
with respect to the partition function (\ref{conto37}).
In the large-$n$ limit we can use the factorisation property: 
$\ev{|\trace U_\Gamma |^2}$ = $\ev{\trace U_\Gamma}$ 
$\ev{\trace U_\Gamma^\dagger}$.
Moreover due to the following symmetry of  the action (\ref{conto37}):
\begin{equation}
\label{joe8}
U_\mu^{(a)} \rightarrow\zeta_\mu^{(a)} U_\mu^{(a)} , \hskip 2cm
\zeta_\mu^{(a)}\in {\bf Z}_n~~,
\end{equation}
which is  unbroken in the weak coupling (large $\beta_H$) regime, we have:
\begin{equation}
\label{joe9}
\ev{\trace U_\Gamma } = \left\{
\begin{array}{ll} 
0 & \mbox{if $\Gamma$ is not a ``tree''} \\
\trace {\bf 1} = n & \mbox{if $\Gamma$ is a ``tree''~~,}
\end{array} \right.
\end{equation}
where by  ``tree" we denote any closed path $\Gamma$
that contains $U_\mu^{(a)}$ and $U_\mu^{(a)\dagger}$ the same number of times
for any $\mu$ and $a$. 
By using this property, eq. (\ref{conto38}) reduces simply to
\begin{equation}
\label{conto39}
\rho_a^2= {1\over  \beta_H}\sum_l {n(l)\over 
(m^2)^{l+1}}~~,
\end{equation}
where $n(l)$ is the number of trees of fixed length $l$ on the lattice
of Fig. 2 starting end ending in the point $a$.
The instability of the weak coupling configuration occurs
at the value of $m^2$ such that the series at the r.h.s. of eq. 
(\ref{conto39}) diverges. 

The basic point now is that\footnote{For related discussions, see
\cite{makeenko}.}, irrespectively of the choice of the lattice 
on which the Kazakov-Migdal model (with quadratic potential) is defined, 
the expectation value of 
$\langle \trace \Phi^2 \rangle$ on a site of the lattice is given by
(\ref{conto39}) where $n(l)$ is the number of trees of fixed length $l$ on the
lattice we considered, starting end ending in the chosen point.
On the other hand, $n(l)$ depends only from the coordination number of the
lattice, and hence it is the same for the lattice of Fig. 2 and for a
$d$-dimensional hypercubic lattice.
In the latter case we know the exact solution,  a semicircular 
distribution of radius given by (\ref{conto26});  $\langle 
\trace \Phi^2 \rangle$ is
proportional to the square of such radius and it becomes singular at the 
value of $\beta_H$ given in eq. (\ref{conto27}).
In conclusion, although we have not been able to solve explicitly for
all $L$'s the model (\ref{nuova33}) without making the explicit assumption
that the $Z_L$ symmetry of the vacuum is unbroken, we have proved that
(\ref{nuova33}) becomes singular at exactly the same critical value of 
$\beta_H$ as our $Z_L$ invariant solution. This shows that no phase 
transition associated to the breaking of the $Z_L$ symmetry
occurs and it is a conclusive evidence
that the $Z_L$ invariance of the vacuum is preserved.
\section{Conclusions}
\label{conclsec}
We have considered the one plaquette matrix model corresponding to a hot 
twisted Eguchi-Kawai model in which the time-like plaquettes are neglected, 
and we have found a solution by calculating the quadratic quantum fluctuations
around the classical vacuum, namely around the twist eating configurations.
The solution for the time-like link variables consists of $L$ bunches of  
eigenvalues, centered around the $L^{\rm th}$ roots of unity, each bunch with a
semicircular distribution of eigenvalues.
This eigenvalue distribution, although rather unusual, does not come as  a 
surprise in this context  because it corresponds to a spreading of 
the eigenvalue distribution of the classical vacuum of the 
Twisted Eguchi-Kawai model.
Our solution becomes unstable at a critical value of the coupling constant,
where a phase transition occurs. The crucial point here is that our calculation
neglects the contributions of the other extrema of the action, so that we do not 
know, without further information, the range of validity of the solution.
This problem is solved by observing that in the hamiltonian limit ($L 
\rightarrow \infty $) the original model is mapped into one with a heat-kernel 
action, which was solved in a previous paper ~\cite{largeN}. As the two 
solution coincide we can safely say that the contribution from the relative
extrema of the action can be neglected until the instability sets in.
We can actually infer that the instability of the solution is determined 
by a condensation of the local extrema of the action. In fact it was shown in 
~\cite{largeN} that the phase transition in the heat-kernel action is
mathematically the same as the Douglas-Kazakov phase transition in QCD2,
which is known to be due to a condensation of classical solutions (instantons).
An interesting possibility is that the full theory, namely the one including the 
space-like plaquettes, might follow exactly the 
same pattern.  It has already been suggested ~\cite{baal} that 
a condensation of the extrema of the classical action is responsible  for
confinement  in the Twisted Eguchi-Kawai model; the results of the present 
paper suggest   that the fluctuations around the twist eating configurations
are the only ones that need to be taken into account above the deconfinement 
transition  and  that they determine the master field of the theory in 
that phase.
An instability of the corresponding solution would then give a lower bound
for the critical coupling, without having to take into account explicitly
the complicated structure of the local extrema, whose condensation is ultimately
responsible of the deconfinement transition at the critical temperature.
The comparison between column {\it (a)} and {\it (c)} of Table I shows that 
the instability of the solution in the deconfined phase and the condensation
of ``instanton" solutions are indeed very close to each other, at least 
in our simplified model.
\vskip 0.5cm
\centerline{{\bf  Acknowledgements}}
\vskip 0.3cm
We thank M. Caselle and S. Panzeri for many helpful discussions.
This work was supported in part by EEC Science program SC$1^*$CT92-0789.
\appendix
\section{Appendix}
\label{appea}
In this Appendix we derive the summation formula given in 
eq. (\ref{conto11}), which is crucial in the computation of the 
quadratic effective model of section \ref{quadmodel}.

This formula can be derived starting from the identity
\begin{equation}
\label{appe3}
\prod_{l=0}^{L-1} 2 \sin (x + {\pi l\over L}) = 2 \sin Lx~~,
\end{equation}
which in turn can be proved by expanding the sine functions in exponentials
and by noticing that the $L^{\rm th}$ roots of unity satisfy the 
property:
\begin{equation}
\label{appe6}
\sum_{k=0}^{L-1} \ee{-{2\pi\ii\over L} kn} = 0\, , \hskip 2cm
n = 1, 2, \dots , L-1.\,\, .
\end{equation}
By taking the logarithm of both sides of eq. (\ref{appe3}) one obtains the
relation:    
\begin{equation}
\label{appe2}
\sum_{l=1}^{L-1} \log \left[2~\sin (x + {\pi l\over L})\right] = 
\log {\sin Lx\over \sin x}~~,
\end{equation}
which is the ``generating identity'' of a set of non-trivial 
trigonometric summation formulas. 
In fact by expanding  the two sides of eq. (\ref{appe2}) in powers of 
$x$ we have:
\begin{equation}
\label{conto41}
\begin{array}{lc}
O(x):\hskip 1.5cm &  \sum_{l=1}^{L-1} \cot {\pi l\over L} = 0~~,\cr
\null & \null \cr
O(x^2): &   \sum_{l=1}^{L-1} \sin^{-2} {\pi l\over L} 
= {L^2 - 1\over 3}~~,\end{array}
\end{equation}
where the last equation coincides with (\ref{conto11}).
It is worth to remark here that,
by taking appropriate combinations of higher order identities, one
can evaluate from the power expansion of (\ref{appe2}) all the sums of the 
form $\sum_{l=1}^{L-1} \sin^{-2k} {\pi l\over L}$.
For instance, form the fourth and sixth order in $x$  one gets:
\begin{equation}
 \sum_{l=1}^{L-1} \sin^{-4} {\pi l\over L}={(L^4 - 1) \over 45}+ {
2 (L^2 - 1) \over 9}
\label{k4}
\end{equation}
and
\begin{equation}
 \sum_{l=1}^{L-1} \sin^{-6} {\pi l\over L}={2 (L^6 - 1) \over 945}+
{(L^4 - 1) \over 45} + {8 (L^2 - 1) \over 45}. 
\label{k6}
\end{equation}



\end{document}